\documentstyle[12pt]{article}
\begin{document}
\pagestyle{empty}
\centerline{               \hfill FERMILAB-PUB-96/180--T}
\centerline{               \hfill NHCU--HEP--96--20}
\centerline{hep-ph/yymmxxx  \hfill UICHEP-TH/96--10}
\vskip 1cm

\begin{center}
{\large \bf Triple Pseudoscalar Decay Mode of Z boson}
\end{center}
\vskip 1cm
\centerline{
Darwin Chang$^{(1,2,3)}$ and
Wai-Yee Keung$^{(2,3)}$ }\vskip 1cm
\centerline{\small\it 
$^{(1)}$Physics Department,
National Tsing-Hua University, Hsinchu, Taiwan, R.O.C.}
\centerline{\small\it 
$^{(2)}$Department of Physics, University of Illinois at Chicago, 
Illinois 60607--7059}
\centerline{\small\it 
$^{(3)}$Fermi National Accelerator Laboratory, P.O. Box 500, Batavia, 
Illinois 60510}
\centerline{(\today)}
\vfill
\begin{abstract}
We analyze the production of triple pseudoscalar Higgs bosons in the
decay channel of $Z\rightarrow AAA$ for light pseudoscalar bosons when the
corresponding scalar boson is too heavy to be produced by $Z$ decay.
Analytic results are obtained both at the tree level and at the one--loop
level.  The branching fraction can be as large as $10^{-5}$ which should 
be detectable at LEP.
\end{abstract}

\vskip 1in

\centerline{PACS numbers: 14.80c , 13.38 , 12.60.F \hfill}
\newpage
%
\pagestyle{plain}
There is essentially no stringent and model--independent limit on the
mass of a pseudoscalar Higgs boson, generically denoted by $A$ in this
paper.  Such a pseudoscalar boson always exists in the extended Higgs
sector beyond the Standard Model (SM).  An identical pair of 
pseudoscalar bosons cannot be
produced in pair in the $Z$ decay, as it is forbidden by Bose statistics. 
The potential limit on the mass of a pseudoscalar Higgs boson comes from
LEP experiments.  However, in all the
analyses\cite{r:pseudo,r:pseudo2,r:haa},
the pseudoscalar bosons are assumed to be produced by the decay of a
physical scalar boson $h$.  For the case when the scalar boson is heavy
(such as $m_h > m_Z$), no limit on $m_A$ has been extracted yet.
If the scalar partner $h$ is heavy enough, the mode $Z\rightarrow hA$ 
will not be allowed by kinematics.  Nevertheless, the channel
$Z\rightarrow AAZ^* \rightarrow AAl^+l^-$ is allowed if $A$ is light
enough.  However, its branching fraction was shown \cite{r:haber} to be
typically about  $10^{-8}$, too small to be detectable for LEP.
A pseudoscalar Higgs boson lighter than a $b$ quark can be ruled out by
$b \rightarrow s A$\cite{r:pseudo3}, however the conclusion will be very
much model dependent.  (Therefore, it is still worthwhile to make a direct
search at LEP even if the pseudoscalar mass is in this light range.)  
In any case, for a pseudoscalar boson
whose mass is heavier than the $b$ quark 
and whose companion scalar boson is too heavy
for the decay $Z\rightarrow hA$, the current model independent bound on
its mass is very weak.   

In this note, we look into another potential discovery channel 
$Z\rightarrow AAA$ for the pseudoscalar boson which may be detectable
among the rare $Z$ decays.  The channel is particularly interesting when 
the lightest scalar is heavier than the lightest pseudoscalar boson which 
can also be an axion.
Experimentally, the $AAA$ final states was searched for by LEP
detectors \cite{r:pseudo} assuming that two $A$'s are decay product of a
physical scalar boson $h$.  A lot of that analysis can probably be
borrowed immediately to the case when $h$ is of-shell.  

In the popular Minimal Supersymmetric Standard Model (MSSM), at tree level
the Higgs masses obey relations\cite{r:pseudo3} 
$m_h < m_Z < m_H$, $m_h < m_A < m_H$ and $m_{H^\pm} > m_W$, where $m_h$ is
defined to be the lighter one of the two scalar bosons.  
These relationships are modified when one loop corrections, due to top
quark, are taken into account.  $m_h$ in this case no longer has to be lighter
than $m_Z$.  However, it is still constrained to be lighter than
about  140 GeV (for $\tan \beta > 1$)\cite{r:mssm}.  
With the radiative corrections,
it is also possible\cite{r:haa} that $m_A < {1\over2}m_h$.
In this sense, our
analysis is also very much relevant to the MSSM in addition to the more
general models.

The one--loop amplitude for $ Z\rightarrow AAA$ via the virtual
top--quark was roughly estimated by Li\cite{r:lfli}.  Here we 
study in details both the tree-level process due to a virtual scalar
Higgs boson, and the one--loop process due to the top--quark loop.

In Fig.~1, we illustrate one of the Feynman diagrams that the triple
pseudoscalar decay mode $Z\rightarrow AAA$ occurs through the gauge
vertex $Z\rightarrow Ah^*$, followed by $h^*\rightarrow AA$\cite{r:haa}.
Phenomenologically, one can describe the interaction among the scalar and 
the pseudoscalar Higgs bosons by an effective Lagrangian, 
\begin{equation}
{\cal L}=\lambda\langle V \rangle hAA \ .
\end{equation}
The coefficient $\lambda\langle V \rangle$ is related to the vacuum
expectation value of the Higgs field, the quartic bosonic couplings and
also some mixing angles. Its value is about the scale of the
electroweak interaction.

The amplitude for 
$Z(p_Z,\varepsilon_Z) \rightarrow A(p_1) + A(p_2) + A(p_3)$ 
can be written in term of the form factors as,
\begin{equation}
{\cal M}=\left[F^h(p_2,p_3)p_1^\nu+F^h(p_1,p_3)p_2^\nu
              +F^h(p_1,p_2)p_3^\nu\right]\cdot (\varepsilon_Z)_\nu
\quad,
\label{e:amp}
\end{equation}
where
\begin{equation}
F^h(p_2,p_3)={g \lambda\langle V \rangle\over\cos\theta_W }
 {2\over (p_2+p_3)^2-M_h^2}
\quad .
\end{equation}
We simplified our picture by assuming that the contribution of the
lightest scalar Higgs boson dominates. 
The analysis is parameterized model independently such that
it would be straight forward to adapt our study to a specific model.

From Eq.~(\ref{e:amp}), we obtain the spin-summed amplitude squared as
follows,
\begin{eqnarray}
\sum|{\cal M}|^2&=&
{4\pi\alpha \lambda^2\langle V\rangle ^2
\over x_W(1-x_W) M_Z^2}
\left[
{2xy-4(1-z-a)\over (1-x+a-h)(1-y+a-h)}+{z^2-4a\over (1-z+a-h)^2}
\right]
\nonumber \\
&\ &
+\quad \hbox{2 other permutations of }x,y,z   \ .
\end{eqnarray}
Here $a=M^2_A/M_Z^2$, $h=M^2_h/M^2_Z$ and 
$x=2p_1\cdot p_Z/M_Z^2$,
$y=2p_2\cdot p_Z/M_Z^2$,
$z=2p_3\cdot p_Z/M_Z^2$, with $x+y+z=2$.
The allowed $x$ and $y$ ranges are
\begin{eqnarray}
2\sqrt{a}        \le &x& \le 1-3a   \ ,\nonumber\\
1-{1\over2}(x+d) \le &y& \le 1-\frac{1}{2}(x-d) \ ,   \\
\hbox{with  }  d      &=& (x^2-4a)^{1\over2}[1-4a/(1-x+a)]^{1\over2} \ .
\end{eqnarray}

The partial width for the channel $Z\rightarrow AAA$ is

\begin{equation}
d\Gamma(Z\rightarrow AAA)= {M_Z\over 256 \pi^3} 
\left({1\over 3}\sum|{\cal M}|^2 \right){dxdy\over 3!}  \ .
\end{equation}
In Fig.~2, we demonstrate the branching fractions of the process
$Z\rightarrow AAA$ for different scenarios, (a) $M_h=90$ GeV (dashed),
(b) $M_h=100$ GeV (solid), and (c) $M_h=110$ GeV (long dashed), for
the case $\lambda\langle V \rangle =100$ GeV.  We find that the size of 
the branching fraction can be as large as $10^{-5}$ for lighter $h$.

Next, we look at the induced amplitude at the one--loop level. This is
potentially significant when $M_h$ is very large so that the tree-level
amplitude is not important.  It is also interesting because it only
depends on the coupling of the pseudoscalar boson with the top quark, and
independent of the details of the Higgs self-couplings.
We shall parameterize the Yukawa coupling
of the top quark to the light pseudoscalar boson ($A$) in the following model
independent form,
\begin{equation}
{\cal L}_{\hbox{\tiny Yukawa}}=g_t \bar t i\gamma_5 t A \quad.
\end{equation}
The coupling between the top quark and the $Z$ gauge boson is given in
the Standard Model,
$$g_Z^t={ e\over 4\sin\theta_W \cos   \theta_W} \quad.
$$
For the top--induced amplitude of the process 
$Z(p_Z)\rightarrow A(p_1)A(p_2)A(p_3)$, there
are six Feynman diagrams. Under charge conjugation, they pair up into 
three sets,
\begin{equation}
{\cal M}=3 {2g_t^3 g_Z^t m_t\over 96\pi^2 }
\int\int\int 3! d\alpha d\beta d\gamma\left(
{   {\cal N}^\nu_{123}  
  \over
      (\mu^2_{123})^2
}
+
{   {\cal N}^\nu_{132}  
  \over
      (\mu^2_{132})^2
}
+
{   {\cal N}^\nu_{213}  
  \over
      (\mu^2_{213})^2
}\right)   \cdot (\varepsilon_Z)_\nu
\quad,
\label{eq:ND}
\end{equation}
\begin{equation}
\hat p_{123}=\gamma p_1-\delta p_3
-\hbox{$1\over2$}(\alpha+\gamma-\beta-\delta)p_Z
\quad,
\end{equation}
\begin{equation}
\mu_{123}^2=
          \hat p_{123}^2
         -\hbox{$1\over4$}M_Z^2
         -(\gamma+\delta)m_A^2
         +m_t^2+(\delta p_3+\gamma p_1)\cdot p_Z
\quad,
\end{equation}
\begin{equation}
{\cal N}^\nu_{123}=
p_2^\nu (M_Z^2-4m_t^2-8\mu^2_{123} + 4 \hat p^2_{123})
+4[(p_3-p_1)^\nu  p_Z -2p_1^\nu p_3+2p_3^\nu p_1]\cdot \hat p_{123}
\quad.
\end{equation}
The Feynman parameters satisfy $\alpha+\beta+\gamma+\delta=1$ and $ 0
\le \alpha,\beta,\gamma,\delta \le 1$.  
In Fig.~3, we carefully show the choice of the momentum flow and the
corresponding Feynman parameters. Our results can be easily produced
following such convention.
A color factor 3 has been explicitly included in Eq.(\ref{eq:ND}).
It is straightforward to generalize expressions (10--12) to other
cases 132 and 213 by permutations.
The charge conjugated diagrams give equal contributions as one can
easily check that ${\cal N}_{123}={\cal N}_{321}$ etc.
To arrange this one--loop amplitude in a similar form as (\ref{e:amp}),
we introduce the form factors $F^t(p_i,p_j)$ in parallel with $F^h$, 
\begin{eqnarray}
F^t(p_1,p_3)=3{2g_t^3 g_Z^t m_t\over 96\pi^2 }
\int\int\int 3!  d\alpha d\beta d\gamma
&&  \hskip -20pt
\left[
 { (4 p_Z + 8 p_1)\cdot \hat p_{132}
\over (\mu_{132}^2)^2}
+{ (4 p_Z + 8 p_3)\cdot \hat p_{312}
\over (\mu_{312}^2)^2}
\right.
\nonumber\\
&&\left.
+{M_Z^2-4m_t^2-8\mu_{123}^2+4\hat p_{123}^2
\over (\mu_{123}^2)^2}
\right]
\end{eqnarray}
The above formulas are ready for numerical integrations. However, for
the purpose of illustration we only extract the leading contribution
in the large $m_t$ limit even though the correction can be of order of
$M_Z^2/m_t^2 \approx 0.25$.  We also remove irrelevant constant terms 
from $F^t$.  Such constant terms cannot contribute to the overall amplitude 
for a physical polarization $\varepsilon$ of the $Z$ boson. We obtain,
\begin{equation}
F^t(p_1,p_3) \approx 3{g_t^3 g_Z^t (p_1\cdot p_3)\over 8 \pi^2 m_t^3}
\quad .
\end{equation}
Unfortunately, such a top-loop induced amplitude is so small that it
produces, by itself, a negligible branching fraction for $Z\rightarrow
AAA$ below 10$^{-10}$, even we assume a SM coupling
$g_t=(\sqrt{2}G_F)^{1\over2} m_t$. 
This is much smaller than the previous rough estimate\cite{r:lfli} by many
orders of magnitude.
More likely, the signal of 
$Z\rightarrow AAA$ comes from the Higgs mediated process.

Since one requires the scalar boson $h$ to be light enough (such as 
90  GeV) in order to get a large branching ratio, one may also consider
the alternative production of $e^+ e^- \rightarrow Z^* \rightarrow hA
\rightarrow AAA$.  This possibility is already
covered in some of the Higgs search analysis\cite{r:pseudo,r:pseudo2}.

As the accumulated events of $Z \rightarrow hadrons$ among the four LEP
groups have reached $10^{7}$, A branching ratio of $10^{-5}$ is
potentially detectable.
The main difficulty seems to be finding a clear signal with high
efficiency for such events.  
If the pseudoscalar boson is heavier than $2 m_b \approx 10$ GeV, then
presumably it will decay dominantly into six $b$ quarks.  
In MSSM\cite{r:guide}, for 
bosons decay into $b \bar{b}$ about $90 \%$ of the time and about $6-8 \%$
into 
$\tau^+ \tau^-$.

For the case 10 GeV $\approx 2 m_b > m_A > 2 m_{\tau} \approx 3.5$ GeV, 
the $A$ boson can decay dominantly
into six $\tau$ leptons or six charm quarks.  The two modes are
competitive with each other.  One can search for 
$\tau^+ \tau^-$ plus four jets or 
$\tau^+ \tau^- \tau^+ \tau^-$ plus two jets or 
$\tau^+ \tau^- \tau^+ \tau^- \tau^+ \tau^-$ in increasing
detection efficiencies.
The answer will depend on the relative fraction between 
$\tau^+ \tau^-$ and $c \bar{c}$ final states.  

For the channel $Z \rightarrow AAA \rightarrow b \bar{b} b \bar{b} b
\bar{b}$, the clear signal can be a number of $b$-tagged jets.
Similar signal was searched before in the previous Higgs search
analysis \cite{r:pseudo} for lighter on-shell scalar boson $h$ using
the same $AAA$ final state.  It was concluded\cite{r:pseudo} that the
current limit of this branching ratio is at about $10^{-4}$ level.
With more recent data, this limit may be improved by a factor of $3$
or more with improved statistics.  To improve this further, one
probably has to increase the efficiency in the identification of six
jets from the three $A$ bosons and the efficiency in b-tagging.
Typically a prize (of about $20-30 \%$) has to be paid to impose tight
cut to reject 3, 4 or 5 jets events.  The efficiency will be higher
for lighter pseudoscalar.  In addition, one has to pay a prize for
b-tagging.  The current b-tagging efficiency of LEP detectors is
roughly about $20\%$ per jet.  Even if one tags only 3 out of 6 jets,
the prize is already quite severe ($20 \times (0.2)^3 = 16\%$).  These
two effects combine to give $3-5 \%$ efficiency of identifying $Z
\rightarrow AAA$. (OPAL\cite{r:pseudo} quoted $6-11 \%$, but with
rather high background).

In a general multi-doublet extension of Standard Model, it is possible
that the pseudoscalar decays into $\tau$ leptons or $b$ quarks with
similar branching fraction, in that case, the best modes to discover
pseudoscalar boson may be $b \bar{b} \tau^+ \tau^- \tau^+ \tau^-$ or
$b \bar{b} b \bar{b} \tau^+ \tau^-$ final states.\cite{r:yhchang} As
far as we know, these modes have not been seriously searched for by
LEP yet.

Our analysis indicates that there is a good chance that one can detect
signal of pseudoscalar boson in $Z$ decay with a branching ratio of
about $10^{-5}$.  It is certainly far from trivial; however, the
reward is that the pseudoscalar boson may be already there in the data
waiting to be uncovered.  
\vfil
\section*{Acknowledgment}
We thank Ling--Fong Li for useful discussion and encouragement for
the completion of this work.  
D.C. also wish to thank Willis Lin, Augustine Chen and especially Yuan-Han
Chang for illuminating discussions on LEP data analysis.  This work was
supported in part by the United States Department of Energy under Grant
Number DE-FG02-84ER40173 and by a grant from National Science Council of 
Republic of China.

\eject

\eject
\topmargin=-0.5in    \headheight=0in       \headsep=0in   \textheight=9.7in
\textwidth=7in       \footheight=2ex 	   \footskip=3ex  \oddsidemargin=-.25in
\evensidemargin=-.25in \hsize=7in          \parskip=0pt   \lineskip=0pt
\abovedisplayskip=3mm plus.3em minus.5em
\belowdisplayskip=3mm plus.3em minus.5em
\abovedisplayshortskip=2mm plus.2em minus.4em
\belowdisplayshortskip=2mm plus.2em minus.4em
		\setlength{\unitlength}{.9mm} 
\section*{Figure Captions}
\begin{enumerate}

\item \label{fig1} One of the tree--level Feynman diagrams for the
process $Z\rightarrow AAA$ via the scalar $h$ Higgs boson. Two other
diagrams are obtained by permuting the momenta.
\item \label{fig2} Predicted branching fractions of the tree--level
process $Z\rightarrow AAA$ for different scenarios, (a) $M_h=90$ GeV
(dashed), (b) $M_h=100$ GeV (solid), and (c) $M_h=110$ GeV (dot-dashed). 
We have set $\langle V\rangle = 100$ GeV.
\item \label{fig3} One of the one--loop Feynman diagrams for the
process $Z\rightarrow AAA$ via the virtual top--quark.  The choice of
momentum flow and the Feynman parameters are clearly labeled.  Two
other diagrams are obtained by permuting the momenta.
\end{enumerate}
\vfill
\begin{figure}[h]
\begin{center}
\begin{picture}(85,70)(15,15)
\thicklines 
\multiput(16,50)(4,0){7}{\oval(2,2)[t]}
\multiput(18,50)(4,0){7}{\oval(2,2)[b]}
\put(30,48){\makebox(0,0){\vector(1,0){5}}}
\put(30,55){\makebox(0,0)[b]{$Z^\nu(p_Z)$}}
\multiput(43,50)(5,5){6}{\line(1,1){4}}
\put(53,60){{\vector(1,1){1}}}
\put(45,58){{$h$}}
\put(68,75){{\vector(1,1){1}}}
\multiput(43,50)(5,-5){6}{\line(1,-1){4}}
\put(58,36){\makebox(0,0){\vector(1,-1){1}}}
\multiput(58,65)(5,-5){3}{\line(1,-1){4.2}}
\put(68,55){{\vector(1,-1){1}}}
\put(85,78){\makebox(0,0)[t]{$A(p_1)$}}
\put(85,48){\makebox(0,0)[t]{$A(p_2)$}}
\put(85,18){\makebox(0,0)[t]{$A(p_3)$}}
\end{picture}
\end{center}
\begin{itemize}
\item[Fig.~1] 
One of the tree--level Feynman diagrams for the process $Z\rightarrow
AAA$ via the scalar $h$ Higgs boson. Two other diagrams are obtained
by permuting the momenta.
\end{itemize}
	\end{figure} 
\eject
\pagestyle{empty}
\begin{figure}
\begin{center}
\begin{picture}(85,70)(15,15)
\thicklines 
\multiput(16,50)(4,0){7}{\oval(2,2)[t]}
\multiput(18,50)(4,0){7}{\oval(2,2)[b]}
\put(30,48){\makebox(0,0){\vector(1,0){5}}}
\put(30,55){\makebox(0,0)[b]{$Z^\nu(p_Z)$}}
\put(43,50){\line(1,1){30}}
\put(58,63){\makebox(0,0){\vector(1,1){1}}}
\put(45,70){{$l+{1\over2}p_Z$}}
\put(50,63){{$\alpha$}}
%
\put(43,50){\line(1,-1){30}}
\put(58,36){\makebox(0,0){\vector(-1,1){1}}}
\put(45,30){{$l-{1\over2}p_Z$}}
\put(50,23){{$\beta$}}
%
\put(73,20){\line(0,1){60}}
\put(73,65){\makebox(0,0){\vector(0,-1){1}}}
\put(75,65){{$l+{1\over2}p_Z-p_1$}}
\put(68,65){{$\gamma$}}
\put(73,35){\makebox(0,0){\vector(0,-1){1}}}
\put(75,35){{$l-{1\over2}p_Z+p_3$}}
\put(68,35){{$\delta$}}
\multiput(73,80)(4,0){7}{\line(1,0){2}}
\put(85,80){{\vector(1,0){5}}}
\put(85,78){\makebox(0,0)[t]{$A(p_1)$}}
\multiput(73,50)(4,0){7}{\line(1,0){2}}
\put(85,50){{\vector(1,0){5}}}
\put(85,48){\makebox(0,0)[t]{$A(p_2)$}}
\multiput(73,20)(4,0){7}{\line(1,0){2}}
\put(85,20){{\vector(1,0){5}}}
\put(85,18){\makebox(0,0)[t]{$A(p_3)$}}

\end{picture}
\end{center}
\begin{itemize}
\item[Fig.~3] 
One of the one--loop Feynman diagrams for the process $Z\rightarrow AAA$ via
the virtual top--quark. The choice of momentum flow and the Feynman
parameters are clearly labeled.  Other diagrams are obtained by
permuting the momenta.
\end{itemize}
\end{figure} 
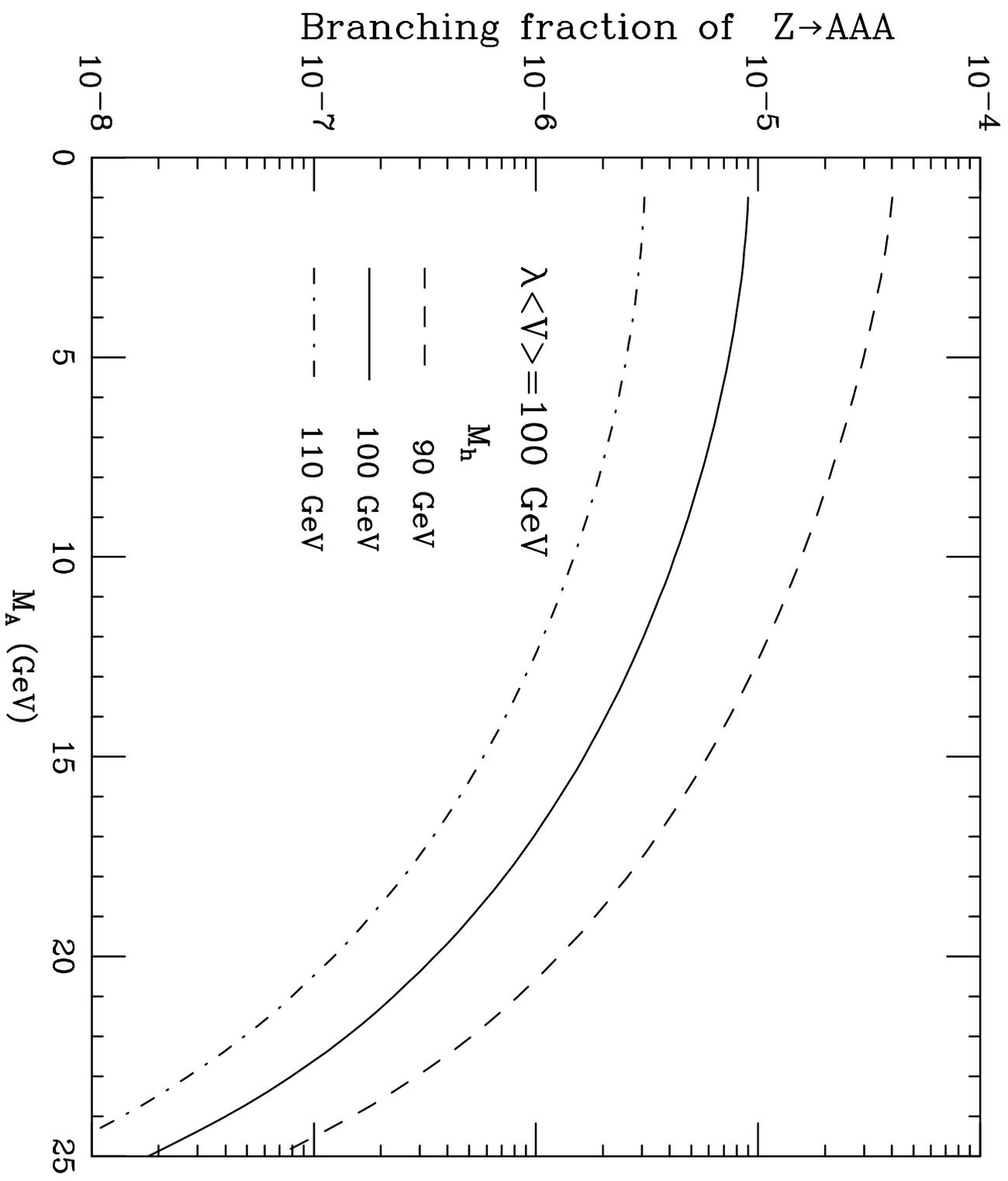
\end{document}